\documentclass[aps,amsmath,prl,twocolumn,a4paper,showpacs,floatfix]{revtex4}

\usepackage{graphicx}

\newcommand{\beq}{\begin{equation}}
\newcommand{\eneq}{\end{equation}}
\newcommand{\bea}{\begin{eqnarray}} 
\newcommand{\enea}{\end{eqnarray}}
\newcommand{\bean}{\begin{eqnarray*}}
\newcommand{\eean}{\end{eqnarray*}}

\begin{document}

\title{Far Infrared absorption of non center of mass modes and optical
  sum rule in a few electron quantum dot with Rashba spin-orbit
  coupling }

\author {P.Lucignano$^{1,2}$} \author{B.Jouault$^3$}
\author{A.Tagliacozzo$^{1,2}$}

\affiliation{$^1$ Coherentia-CNR,  
Monte S.Angelo - via Cintia, I-80126 Napoli, Italy }
\affiliation{$^2$ Dipartimento di Scienze Fisiche Universit\`a di 
         Napoli, "Federico II ", Napoli, Italy}
\affiliation{$^3$ GES, UMR 5650, Universit\'e Montpellier $II$
         34095 Montpellier Cedex 5, France}
\date{\today}
\begin{abstract}
Spin-orbit interaction in a quantum dot couples far infrared radiation
to non center of mass excitation modes, even for parabolic confinement
and dipole approximation. The intensities of the absorption peaks
satisfy the optical sum rule, giving direct information on the total
number of electrons inside the dot.  In the case of a circularly
polarized radiation the sum rule is insensitive to the strength of a
Rashba spin-orbit coupling due to an electric field orthogonal to the
dot plane, but not to other sources of spin-orbit interaction, thus
allowing to discriminate between the two.
\end{abstract}
\maketitle

\textit{Introduction}.

Semiconducting Quantum Dots (QD) with few confined electrons are
possible candidates for applications in future quantum
electronics\cite{kouwenhoven}.  The separation of the electron levels
in such artificial atoms, of size of hundred of nanometers, is of the
order of few $meV$, so that optical spectroscopy requires Far Infrared
Radiation (FIR).  Indeed, FIR absorption is a common tool in large
scale QD arrays\cite{sikorski} (e.g. in $In$ QDs\cite{fricke} or
field-effect confined $GaAs$ QDs\cite{krahne}) ever since their first
fabrication. However, in parabolically confined dots, it is well
established that the FIR spectrum is rather poor of information,
because the dipole approximation holds to a high degree of accuracy
and light couples only to the electron center of mass (CM) modes (so
called Kohn's modes $\omega_{\pm}$). The latter, are free-oscillator
like and are decoupled from the internal dynamics of the correlated
electrons (Kohn's theorem \cite{maksym}). Hence, the location of the
absorption peaks does not depend on the number of electrons $N$
confined in the dot.  Non parabolic corrections to the confinement
potential have been invoked to defeat Kohn's theorem \cite{sikorski},
what would provide also information about $e-e$ correlations by using
FIR.  Indeed, weak plasma modes have been spotted just below the upper
Kohn frequency $\omega_{+}$ \cite{krahne}. Recently, Raman scattering
is improving as a tool to probe correlation effects\cite{pellegrini},
and to prepare spin states in QDs \cite{imamoglu}.

Much interest is being focused on the Rashba Spin Orbit (RSO)
interaction \cite{rashba}, which arises in QD structures due to the
two-dimensional (2D) confinement, since it can be tuned by gate
voltages parallel to the $x-y$ structure\cite{nitta}.  RSO interaction
offers a precious tool for the manipulation of the dot spin states and
is quite relevant for the proposed application of QDs as spin qubits
\cite{divincenzo}. In this context, the study of the effects of the
various spin-orbit (SO) couplings (e.g.  including Dresselhaus, etc.) is
quite crucial, because, in conjunction with the electron-phonon
interaction, it is one of the causes of spin relaxation and dephasing,
which limits the coherent evolution of the
spin\cite{kastner,divincenzo}.  Besides, SO affects
conductivity directly, by turning weak localization corrections into
antilocalization ones, as probed in large QDs\cite{miller}.

In the presence of SO coupling, the CM dynamics of the electrons  and
that of their relative coordinates cannot be
separated\cite{jacak}. Hence a simple FIR experiment on a QD with SO
coupling can probe any excitation mode and correlation effects
compatible with optical selection rules.  By exciting a few electron
QD in the presence of RSO coupling and of an external orthogonal
magnetic field $B$ with circularly polarized light, it is possible to
identify the collective spin excitations, as we discussed
previously\cite{noi,noi2}. The intensities of the $non-$CM peaks
increase with SO coupling at the expenses of the CM ones. We find that
the absorption intensities satisfy the optical sum rule, encountered
in single atoms as well as in solids\cite{pines}.  In particular, the
total intensity is proportional to the number of electrons $N$
confined in the dot. In the case of a circularly polarized light, the
sum rule is independent of the strength of the RSO coupling, even in a
constant magnetic field. This property of the sum rule, that we proof
analytically and check numerically, could help in identifying the
relative weight of the RSO coupling, with respect to other sources of
intrinsic SO interactions \cite{jusserand}.

\begin{figure*}
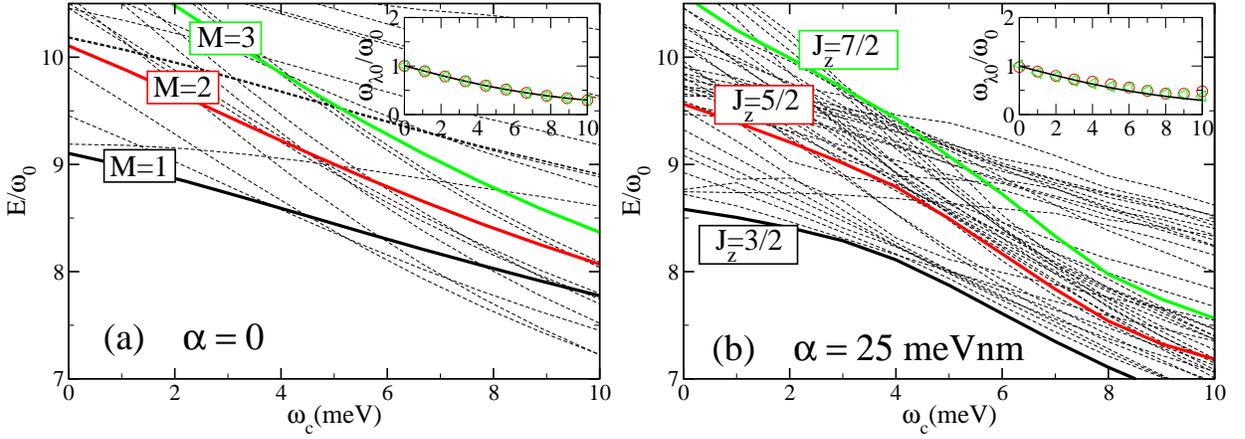

\includegraphics*[width=0.45\linewidth]{prova3cm.eps}
\includegraphics*[width=0.45\linewidth]{lev3so0.4_benv2.eps}
\caption{(Color on line) Low lying energy spectrum for a 3 electron QD
as a function of $\omega_c$ in the absence of RSO $\alpha =0 $({\sl
left panels}) and with RSO $\alpha =25 meV\: nm $ ({\sl right panels})
($\omega_d=5meV$, $U=13meV$). {\sl Main panels}: the low lying levels
are thin dashed lines, the GS and the first two CM excitations
($S=1/2$) are singled out of the plotted levels with heavy lines of
different colors. The total $M$ ({\sl left}) or $J_z$ ({\sl right})
are indicated. {\sl Inset:} the energy difference between successive
CM excitation energies $vs. \: \omega _c $ showing equal spacing,
which are equal to $\omega_-$ (plotted as a black curve) for $\alpha
=0$, but not for $\alpha \neq 0 $. }
\label{lev3}
\end{figure*}

\textit{Center of mass excitations in Quantum Dot}.
The electrons are confined in the $(x,y)$ plane by a parabolic
potential of characteristic frequency $\omega _d $, in the presence of
an uniform orthogonal magnetic field $\vec B = -B \hat z$ .  The total
Hamiltonian for electrons of charge $-e$, interacting via the Coulomb
potential, in the effective mass   ($m_e^*$) approximation, in the
absence of SO coupling, is:
\beq
H   =  \sum _i^N \left [ \frac{1}{2m_e^*} \left(\vec p_i + \frac{e}{c} \vec A_i
\right)^2+\frac{1}{2} m_e^* \omega_d^2 \vec r _i ^2 \right ] 
+ \sum_{ { i,j=1} \atop {i<j}}^{N}
\frac{U}{|\vec r _i - \vec r_j|}
\;\;,
 \label{Dhamiltonian}
\eneq
with $\vec A_i=B/2(y_i,-x_i,0)$. The strength of the Coulomb
interaction $U $ is dictated by the screening in the dot and is a
parameter in our calculation. In the presence of an orthogonal
magnetic field $B$ (cyclotron frequency $\omega _c = eB/m_e^*c$), the
characteristic length due to the the lateral geometrical confinement
$l=\sqrt{\hbar / m_e^*\omega_o}$ depends on the frequency $\omega_o =
\sqrt { \omega_d^2 + {\omega_c^2}/{4}}$.

The ratio $v=U/\omega_o$ gives an estimate of how strong the
correlations are. By tuning $U$ and $\omega_o$, one can range from a
Fermi liquid behavior ($v\sim 1$), up to very strongly correlated
regimes ($v\sim5-7$), in which cristallized phases appear, the so
called Wigner molecule \cite{grabert}.  We will focus on intermediate
regimes $v\sim 2-4$ in what follows, in which correlations are too
strong to be dealt with by using an Hartree Fock approach, but not
enough to allow for breaking of azimuthal symmetry and for creation of
a Wigner molecule\cite{reale}.  We use exact numerical diagonalization
as done previously \cite{jouault}.

The Hamiltonian of Eq.~(\ref{Dhamiltonian}) separates
as: $H = H_{CM} + H_{r}$, with $\left [ H_{CM}, H_{r} \right ] =0$.
Here $H_{CM} $ is the Hamiltonian for the CM coordinates $ \vec R =
\frac{1}{N} \: \sum _i \vec{r}_i$, $\vec P = \: \sum _i \vec{p}_i$,
with total mass $M^*= Nm^*_e $ and $ H _r $ involves only the relative
coordinates $\vec{p}_{ij} \equiv \vec{p}_i - \vec{p}_j$, $\vec{r}_{ij}
\equiv \vec{r}_i - \vec{r}_j$.

Let us first consider the case of $N=2$ electrons for sake of
illustration. The quantum numbers labeling the two particle states are
particularly simple. Indeed, the spin wavefunction factorizes, as well
as the  two-dimensional harmonic oscillator wavefunction of the CM. Finally,
just a single particle orbital wavefunction for the relative
coordinate is present.  The CM wavefunction is always symmetric w.r.to
the exchange of the two particles. Hence, the requirement of overall
antisymmetry for the total wavefunction, fixes reciprocally the
symmetries of the relative motion and of the spin wavefunctions. The
ground state (GS), at zero magnetic field, is a singlet of energy
$E_0=4.097\omega_o$, for $\omega_o=5meV,U=13meV$.  The CM
excitations are always equally spaced at each $B$. Indeed, we find, at
$B=0$, the first two CM excitations at energies $E_1=5.099\omega_o$
and $E_2=6.100 \omega_o$ with a relative error $\sim2/1000$ w.r. to
the correct result.

\begin{figure*}
\includegraphics[width=0.4\textwidth]{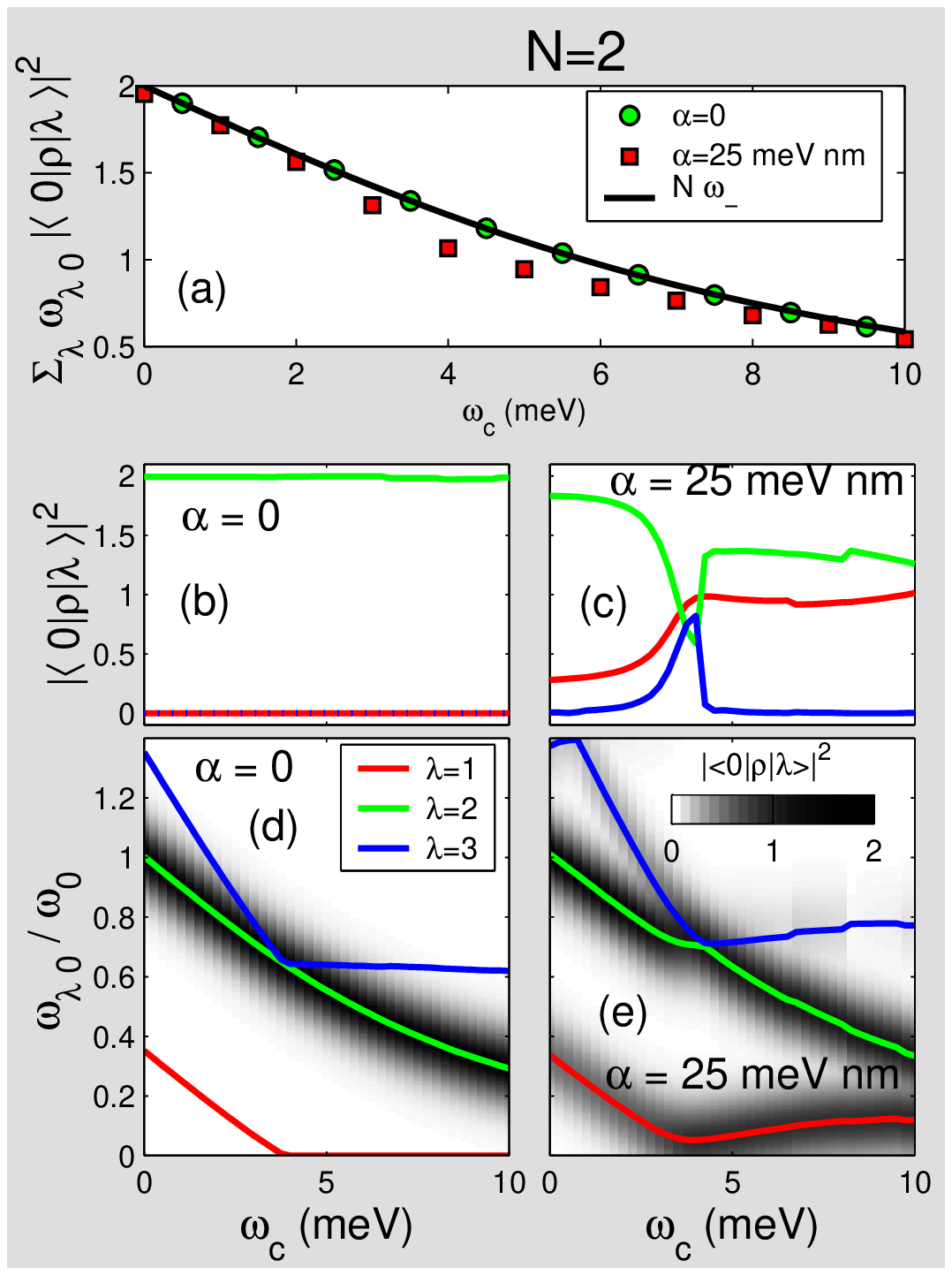} \hspace{0.1\columnwidth}
\includegraphics[width=0.4\textwidth]{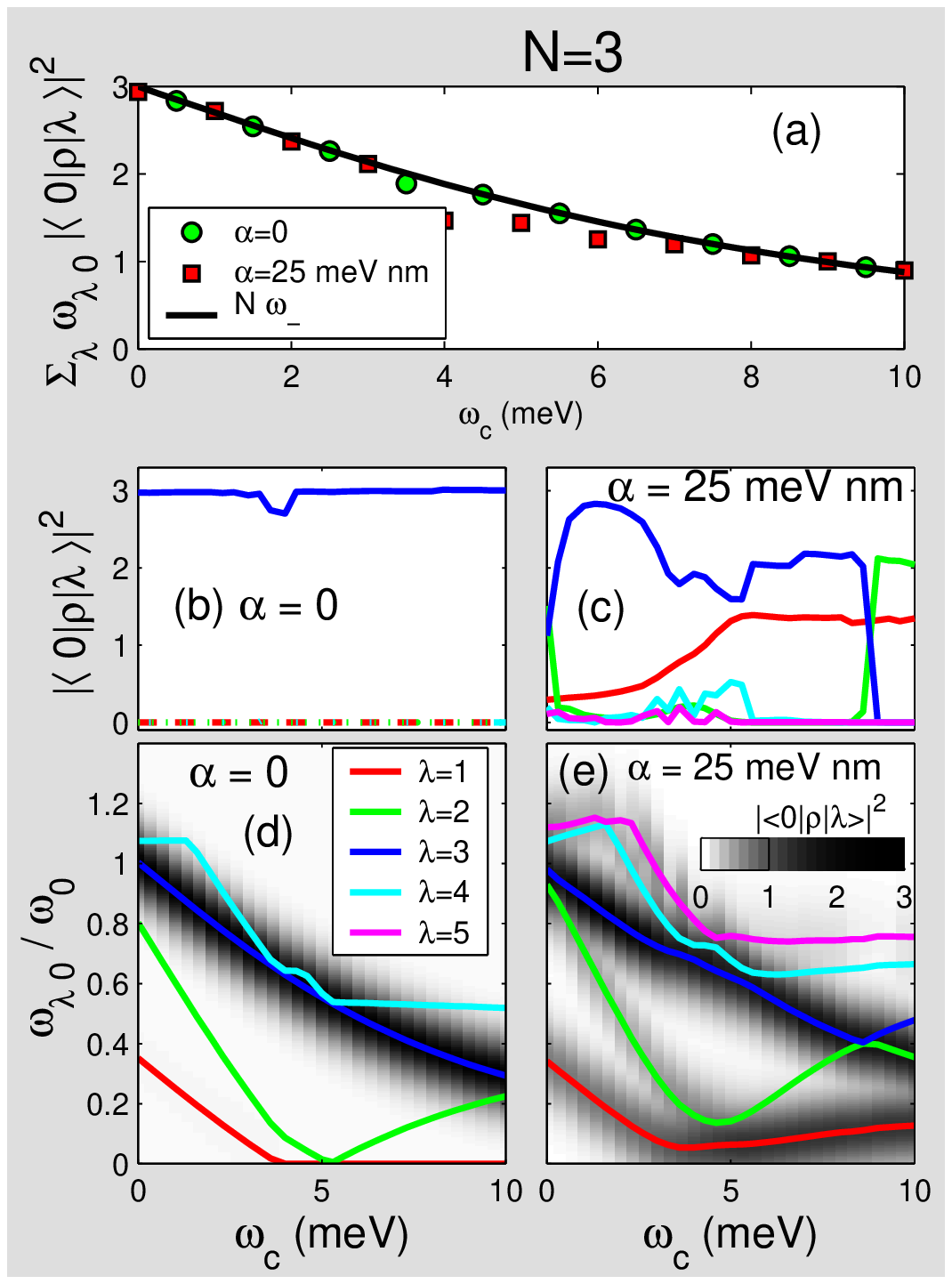}
\caption{(Color online) N=2 ({\sl left}) and N=3 ({\sl right}). {\sl
Bottom panels}: the excitation modes energies $\omega_{\lambda 0}
\equiv E_{\lambda} -E_{GS} $ $ vs. \: \omega_c $, in units of
$\omega_o$, for $\alpha = 0$ (d) and $ 25 meV\: nm$ (e). The
intensities of their FIR absorption peaks are represented by a
greyscale plot underlying the heavy lines.  At $\alpha =0 $ a single
peak appears, corresponding to the CM excitation mode.  By increasing
the RSO, other excitation branches appear, whose weight increases with
magnetic field. The broadening is artificial and equal for all the
peaks \cite{binota}. {\sl Middle panels:} the intensities of the
excitation modes are plotted with colors corresponding to the curves
in {\sl (d),(e)}, in arbitrary units.  The top plots show the sum rule
vs $\omega_c$ in units of $\omega_o$. The plotted sum is independent
of the $RSO$ interaction. Numerical discrepancy occurs in the
neighborhood of a level crossing ( $\omega _c \sim 4 \: meV$ ), due to
the truncation of the Hilbert space. }\label{ap1}
\end{figure*}
 
For more than two electrons the analytical factorization of spin and
orbital wavefunction is possible only when the dot is fully spin
polarized. This implies that, in the general case, the CM modes can be
identified only by comparing the energies of the states
numerically. In Fig.(\ref{lev3}.{\sl a}) we have plotted a few low
lying energy levels for $N=3$ vs $\omega _c$, in the absence of SO
coupling.  Good quantum numbers are the total orbital angular momentum
$M$ orthogonal to the dot disk, the total spin $S$ and the projection
of the spin along $\hat z$, $S_z$.  We have singled out the GS (bold
black line) and the first two CM modes $ E_1, E_2 $ (red and green
line respectively) with the same $S=S_z=0$ as the GS, but with $M$
increasing by one. They are equally spaced at each $B$ (within
numerical errors), as shown in the inset, where the energy differences
$E_1 -E_{GS}$ (red circles), $E_2- E_1$ (green triangles) and the
expected difference $\omega_-$ (black line) are reported in units of
$\omega _o$.

The RSO coupling, included in Fig.(\ref{lev3}.{\sl b}), adds to the non
interacting Hamiltonian of eq.(\ref{Dhamiltonian}) the potential:
\begin{eqnarray}
V^{RSO} = \frac{\alpha}{\hbar} \hat {z} \times \sum _i^N \: \left(\vec
 p_i+\frac{e}{c}\vec A _i\right) \cdot \vec{\sigma_i}\: .
\label{sorbita}
\end{eqnarray}
Here $\vec{\sigma}$ are the Pauli matrices and $\alpha$ is
proportional to the effective (crystal plus applied) electric field in
the $\hat z $ direction. In a biased dot, the actual size of this
perturbation would depend also on the screening of the source drain
bias voltage $V_{sd}$ applied to the contacts. For a reference dot
with $\omega_d = 5meV$ and $ U =13meV $, we choose $\alpha $ in the
range $0\div 25 \: meV \: nm$.  In the presence of
the term of eq.~(\ref{sorbita}) in  the Hamiltonian, 
the orbital angular momentum $M$ and the spin $S_z
$ cease to be separately conserved, while the total angular momentum
along $z$, $J_z=M+S_z$, is.  The total spin $S$ of the CM levels in
Fig.(\ref{lev3}.{\sl b}) keeps being the same as that of the GS.
Nevertheless, $S$ of the GS increases with $\omega _c$ 
due to crossings.

\textit{FIR absorption, Kohn modes, and optical sum rule}.
The FIR interaction in the dipole approximation can be written as
\beq
H_{FIR}= {\mathcal{A}}_0(\omega)\hat\epsilon\cdot
 \vec P \:e^{i \omega t} + h.c.
\eneq
where $A_0(\omega)$ is the envelope function of the FIR wavepacket in
the $\omega$-space, which we suppose to be almost monochromatic and
$\hat\epsilon$ is the polarization vector of the light.  The Kohn
modes are excited at frequency $\omega_{\pm}=\omega_o \pm\omega_c/2$.
It is apparent from eq.(\ref{sorbita}) that the CM and the relative
coordinates no longer decouple, in the presence of
the RSO term. As a consequence, more excitations appear in the FIR
spectrum  and we now discuss the relative intensities of 
the peaks we find.

We choose circularly polarized light in the $x,y$ plane, in order to
excite modes in subspaces with $\Delta J_z=\pm 1$. If we take right
hand polarization $\hat{\epsilon}_R=\hat x + i\hat y$, the
radiation transfers one unity of angular momentum to the dot, $H_{FIR}
\propto \rho_{+}^\dagger +h.c.$ where:
\beq
\rho^\dagger_{+}=
\sum^{'}_{nm\sigma} \frac{m}{|m|} \left(c^\dagger_{n-1m+1\sigma}
c_{nm\sigma}+c^\dagger_{n+1m+1\sigma}c_{nm\sigma}\right)\:\: .
\label{exc}
\eneq
The operators $c_{nm\sigma}$ correspond to the single particle
Darwin-Fock orbitals $\phi _{nm}$ \cite{noi}. These are the
eigenfunctions of the 2D harmonic oscillator with frequency $\omega
_o$ and energy: $ \epsilon _{n ,m} = (n +1) \hbar \omega _o -
\frac{m}{2} \hbar \omega _c \:\:$.  Here $m$ is the angular momentum
in the $z$ direction ( $m \in (-n,-n+2,...,n-2,n)$ with $n \in
(0,1,2,3,...)$ ) and $\sigma $ is the spin projection along the $\hat
z $ axis.
The prime in eq.(\ref{exc}) restricts the sum in such a way that the
labels containing $n$ and $m$ are compatible with the given rules.
$\rho^\dagger_{+}$ is the operator creating an excitation which
increases $M = \sum _i^N m_i $ by one.
The energy  location of the FIR absorption peaks 
 for two ({\sl left panels}) and three electrons ({\sl
right panels}) is  shown   in Fig.(\ref{ap1}) $vs. \: \omega _c $
 for $\alpha = 0$ {\sl (d)}, and
$\alpha= 25 \: meV \: nm$  {\sl (e)} and  their  intensity,
\beq
I_\lambda= |\langle J^{GS}_z+1,\lambda|H_{FIR}|J_z^{GS}\rangle|^2\;,
\eneq
appears as a grey scale plot in arbitrary units.  The broadening of
the lineshapes is artificial and equal for all the peaks.
The excitation energies $\omega_{\lambda 0} \equiv E_{\lambda} -E_{GS}
$, for $\Delta J_z =1 $, are also plotted in units of $\omega _o$
vs. the magnetic field $\propto \omega_c$ as colored curves in
Fig.(\ref{ap1}{\sl .d,e}).  $\lambda $ is a generic label
\cite{binota}.  In the absence of RSO, only the lower Kohn mode
$\omega_{-}$ can be excited for right hand polarization.  By
increasing the RSO coupling, new possible excitations appear below and
above the $\omega_{-}$ mode. In particular the one below increases
markedly in intensity when the magnetic field increases.  We have
shown that it goes almost soft at the crossover to the fully spin
polarized state for the dot \cite{noi,noi2}. This is the collective
spin excitation which recalls the skyrmion ($\Delta S =0$) of a
Ferromagnetic Quantum Hall disk at filling close to one.  Inspection
of the intensities in Fig.(\ref{ap1}{\sl .b,c}) shows that, by
increasing the RSO, the intensity of the Kohn mode drops down and the
lost spectral weight is transferred to the newly emergent modes (the
colors correspond to the $\omega _ {\lambda 0}$ reported in
Fig.(\ref{ap1}.{\sl d,e})).  In fact, we have verified that the
optical sum rule holds, as customary in atoms and solids \cite{pines}.
The optical sum rule, in this case, reads:
\beq
\sum_\lambda \: \omega _{\lambda 0}\: 
 \left|\langle \lambda \left|\rho_{+}^\dagger \right|0\rangle\right|^2=
 \langle 0 \left|\left[\rho_{+}^\dagger , \left[H,\rho_{+}\right]\right]
\right|0\rangle = N  \omega _-
\label{sumrule}
\eneq
where $|0\rangle $ is the GS.  The sum rule does not depend on the
interaction term, nor, remarkably, on the RSO, because
$[\rho_{+}^\dagger , [V^{RSO},\rho_+] ]=0$. At zero magnetic field
$\omega_-=\omega_d$, so that Eq.~\ref{sumrule} sums up to $N\:\omega_d
$ only.  This can be easily checked analytically for non interacting
particles, by using a single Slater determinant for $|0\rangle $.
The sum rule is plotted vs. $\omega _c$ in the top panel of
Fig.(\ref{ap1}{\sl .a, left}) ($N=2$) and Fig.(\ref{ap1}{\sl .a,
right}) ($N=3$). Only few terms were included in the summation over
$\lambda $, because strict selection rules limit the number of states
contributing, up to a relatively high energy.  The curves for $ \alpha
= 0$ and $25 \: meV \: nm $ almost coincide, except for the
neighborhood of $\omega_c \sim 4 \: meV$, where levels cross and a
larger computational Hilbert space in the calculations would be
desirable. Of course, only the CM transition contributes for $\alpha
=0$.

In summary, the inclusion of a spin orbit interaction invalidates
Kohn's theorem for FIR absorption in the dipole approximation, even
for parabolically confined dots.  The possibility to excite non-center
of mass modes makes the optical response of QDs more rewarding.  As in
atoms and solids, the total sum of the oscillator strengths is
proportional to the number of particles in the dot. The optical sum
rule is a precious tool to extract valuable information on correlation
effects in QDs by using the FIR spectroscopy. Here we have considered
a top gate controlled RSO coupling and circularly polarized radiation.
While the relative weight of the intensities of the absorption peaks
depends on the strength of the Rashba spin-orbit coupling, the total
sum rule for the circularly polarized light is insensitive to it. We
suggest that this fact allows to discriminate the amount of SO
coupling present, other than the RSO term $V^{RSO}$ given by
Eq.(\ref{sorbita}).

{\sl Acknowledgements }

We gratefully acknowledge enlightening remarks by B.Jusserand and the
hospitality at the ICTP (Trieste) where this paper was partially
written. Work partly funded by the Italian Ministry of Education and
by ESF FONE project SPINTRA.


\begin{thebibliography}{40}
\bibitem{reale} $GaAs$ and $InAs$ QDs may fall in the chosen range,
 depending on the size. In fact, for a typical confinement energy
 $\omega_d=5meV$, we have $v\sim3$ for $GaAs$ and $v\sim 2$ for $InAs$.
\bibitem{binota} At $B=0$( $\alpha \neq 0$) we find that the total
 spin differences of the excitations, $\Delta S _{\lambda 0}$ are $
 1,0,0$ for $\lambda =1,2,3 $, respectively ($N=2$) and $0,1,0,1$
 for $\lambda =1,2,3,4 $, respectively ($N=3$). At larger $\omega_c
 > 4 meV$, $\Delta S _{\lambda 0}$ changes because a fully spin
 polarized GS sets in.
\end{thebibliography}
\end{document}